\documentclass[a4paper,12pt]{article}
\usepackage{amsthm,graphicx}
\usepackage{a4}
\usepackage{epsfig}
\usepackage{latexsym, amssymb} 
\usepackage{amsmath}
\usepackage{psfrag}
\usepackage{bm}
\usepackage{rotating}
\usepackage{hyperref}
\usepackage[utf8]{inputenc}
\hypersetup{
    unicode=false,          
    pdftoolbar=true,        
    pdfmenubar=true,        
    pdffitwindow=false,     
    pdfstartview={FitH},    
    pdftitle={My title},    
    pdfauthor={Author},     
    pdfsubject={Subject},   
    pdfcreator={Creator},   
    pdfproducer={Producer}, 
    pdfkeywords={keyword1} {key2} {key3}, 
    pdfnewwindow=true,      
    colorlinks=true,        
    linkcolor=red,          
    citecolor=blue,         
    filecolor=magenta,      
    urlcolor=green,         
    linktocpage=true
}

\def\beq{\begin{equation}}
\def\eeq{\end{equation}}
\def\br{\begin{eqnarray}}
\def\er{\end{eqnarray}}
\def\benu{\begin{enumerate}}
\def\efnu{\end{enumerate}}

\def\cl{C_{\ell}}

\def\thefootnote{\fnsymbol{footnote}}

\begin{document}
\title{Test of consistency between Planck and WMAP}
\date{}
\maketitle
\begin{center}
\author{Dhiraj Kumar Hazra}$^{a}$\footnote{dhiraj@apctp.org} and  \author{Arman Shafieloo}$^{a,b}$\footnote{arman@apctp.org}\\ 
%
{\it $^{a}$Asia Pacific Center for Theoretical Physics, Pohang, Gyeongbuk 790-784, Korea\\
$^{b}$Department of Physics, POSTECH, Pohang, Gyeongbuk 790-784, Korea}\\
\end{center}
\begin{abstract} 
Within the context of the concordance model of cosmology we test the consistency of the angular power spectrum data from WMAP and Planck looking for possible systematics. 
The best fit concordance model to each observation is used as a mean function along with a Crossing function with an orthogonal basis to fit the data 
from the other observation searching for any possible deviation. We report that allowing an overall amplitude shift in the observed angular power spectra of the 
two observations, the best fit mean function from Planck data is consistent with WMAP 9 year data but the best fit mean function
generated from WMAP-9 data is not consistent with Planck data at the $3\sigma$ level. This is an expected result when there is no 
clear systematic/tension between two observations and one of them has a considerably higher precision. We conclude that there is no clear
tension between Planck and WMAP 9 year angular power spectrum data from a statistical point of view (allowing the overall amplitude shift). Our result highlights the fact that while the angular power
spectrum from cosmic microwave background observations is a function of various cosmological parameters, comparing individual parameters might be misleading in
the presence of cosmographic degeneracies. Another main result of our analysis is the importance of the overall amplitudes of 
the observed spectra from Planck and WMAP observations. Fixing the amplitudes at their reported values results in an unresolvable tension between the two 
observations at more than $3\sigma$ level which can be a hint towards a serious systematic.

\end{abstract}

\newpage
\tableofcontents
\renewcommand{\thefootnote}{\arabic{footnote}}
\setcounter{footnote}{0}
\section{Introduction}
 Cosmological observations, in particular the data from cosmic microwave background (CMB), have been providing us 
  knowledge of the history and constituents of the Universe since the Cosmic Background Explorer 
 survey. Over time we have been able to constrain our knowledge of the early Universe with increasing precision. Two full sky
 satellite surveys of CMB, the Wilkinson Microwave Anisotropy Probe (WMAP)~\cite{Hinshaw:2012fq} and Planck~\cite{Planck:2013kta,Ade:2013zuv}, have 
 released their data very recently (last release of WMAP data and first release of Planck data). As the most precise CMB observation, Planck has modified
 the derived cosmological parameters that we had obtained from the WMAP and other CMB surveys including the Atacama Cosmology Telescope~\cite{ACT} and the
 South Pole Telescope~\cite{SPT}. Surprisingly, within the same framework of the standard concordance $\Lambda$CDM model, we find that the values of 
 some of the key cosmological parameters derived from Planck data are significantly different than the derived values from earlier CMB observations. For example,
 the values of the Hubble parameter $H_0$ and the dark energy density are found to be considerably less from Planck data compared to the values reported by WMAP. 
 Two questions immediately arise from these differences : first, whether Planck or earlier CMB observations have some unknown systematic that is reflected in their 
reported angular power spectra and second, whether the standard concordance $\Lambda$CDM model is a suitable and flexible model to explain different CMB data 
for large multipoles. In our two paper series we will try to address these two questions.  We address the
consistency of the concordance model of cosmology with Planck data in the other paper of this series~\cite{Hazra:2014hma}. In this paper
we analyze and compare the reported angular power spectra from WMAP and Planck surveys, to search for any significant deviation.

We should note that looking at individual cosmological parameters cannot trivially imply inconsistency between the
two observations. This is basically due to the fact that the standard six cosmological parameters of the concordance model are highly degenerate and not 
orthogonal. In this paper we use Crossing statistic and its Bayesian interpretation~\cite{Shafieloo:2010xm,Shafieloo:2012jb,Shafieloo:2012yh,Shafieloo:2012pm} 
to compare the two datasets in the context of the standard model and independent 
of the individual cosmological parameters. The best fit model to each observation is used as a mean function along with a Crossing function to fit the data from the other observation. 
This allows different smooth variations around a given mean function, allowing us to check whether we can improve the fit to the other data. We have used Chebyshev polynomials as the
Crossing function, as used before in a different context~\cite{Shafieloo:2012pm}. In fact Chebyshev polynomials have properties of 
orthogonality and convergence which make them appropriate as a Crossing function for smooth mean functions. Using the Crossing statistic, the consistency 
of the two datasets can be checked and it can be addressed whether, between the datasets, there lies unknown systematic effects. This paper is organized as follows.
In section~\ref{sec:formalism} we shall describe in detail the framework of the comparison using the Crossing statistic. In section~\ref{sec:results} we provide our 
results and sketch the conclusions.

\section{Formalism}~\label{sec:formalism}
 
 In this section we shall briefly discuss the Crossing statistic and how we use the method to compare two observations. 
 The Crossing statistic was first proposed in~\cite{Shafieloo:2010xm} followed by its Bayesian interpretation~\cite{Shafieloo:2012jb} and was 
 subsequently used in,~\cite{Shafieloo:2012yh,Shafieloo:2012pm} for reconstruction of the expansion history and in searching for systematics in
 data from supernovae and galaxy clusters. The main idea behind the Crossing statistic is that given data based on an actual 
 fiducial model and taking a proposed model, the actual model (hence the data) and the proposed model will cross each other at $1-N$ points. 
 In the Bayesian interpretation of the Crossing statistic one can argue that two different models become virtually indistinguishable if one of them 
 is multiplied by a suitable function. The coefficients of this function constitute the Crossing hyperparameters and the functions themselves 
will be called Crossing functions following~\cite{Shafieloo:2012jb}. A Bayesian interpretation of the Crossing statistic can be used to test consistency of a proposed model 
and a given dataset without comparing the proposed model to any other model. In~\cite{Shafieloo:2012pm} the Crossing statistic has been used to compare two different datasets
, searching for possible systematics, and in this paper we will follow a similar approach. Similar to~\cite{Shafieloo:2012pm} we use Chebyshev polynomials of different orders 
as Crossing functions and we multiply them to a given mean function to fit a dataset. If the given mean function is an appropriate choice to express the data, the 
Crossing hyperparameters (coefficients of the Chebyshev polynomials) would all be consistent with their fiducial values. This basically means that the given mean function does not need any 
significant modification to fit the data better. However, if the best fit derived Crossing hyperparameters deviate significantly from zero, then one can conclude that the 
given mean 
function does not express the data well and including some modifications from the Crossing function will improve the fit to the data significantly. The 
power and accuracy of the method has been shown in previous publications - it can be used for various purposes including regression and searching for systematics. \\

In this paper we consider two datasets~\footnote{Here, by CMB dataset we refer to the derived angular power spectrum from the corresponding survey. On the map 
level the coherence of WMAP-9 and Planck data were investigated in the following paper~\cite{Kovacs:2013vja}.}, 
namely WMAP 9 year and Planck CMB data and we perform our analysis in the framework of the standard $\Lambda$CDM model as a pool of mean functions. 
To test the consistency of the two datasets our algorithm is as follows:\\ 

1. First we fit one of the data sets, let us say Planck data assuming the standard concordance model. We call this best fit model $C_{\ell}^{\rm TT}\mid_{\rm best~ fit~ Planck}$.\\ 

2. Then we assume a Crossing function and multiply it by $C_{\ell}^{\rm TT}\mid_{\rm best~ fit~ Planck}$ to get $ C_{\ell}^{\rm TT}\mid_{\rm modified}$,\\

\beq
C_{\ell}^{\rm TT}\mid_{\rm modified}^{N} =C_{\ell}^{\rm TT}\mid_{\rm best~ fit~ model}~\times ~T_{\rm N}(C_0,C_1,C_2,...,C_N,\ell).
\label{eq:mainmain}
\eeq

In this work we use Chebyshev polynomials of different orders as Crossing functions,\\

\beq
T_{\rm II}(C_0,C_1,C_2,x)=C_0+C_1~x+C_2(2x^2-1)~\label{eq:Crossing-function2}
\eeq

\beq
T_{\rm III}(C_0,C_1,C_2,C_3,x)=C_0+C_1~x+C_2(2x^2-1)+C_3(4x^3-3x)~\label{eq:Crossing-function3}
\eeq

\beq
T_{\rm IV}(C_0,C_1,C_2,C_3,C_4,x)=C_0+C_1~x+C_2(2x^2-1)+C_3(4x^3-3x)+C_4(8x^4-8x^2+1)~\label{eq:Crossing-function}
\eeq

where in our case, $x=\ell/\ell_{\rm max}$ and $\ell_{\rm max}$ is the maximum multipole moment covered by the survey (the maximum multipole covered by whichever 
survey that has the broader range). To have an idea how Crossing functions perform, one can look at the effects of different hyperparameters. Considering Chebyshev 
polynomials as the Crossing function, $C_0$ clearly affects the overall amplitudes and shifts the mean function up and down. $C_1$ adds a tilt to the mean function and 
$C_2$ and higher order coefficients allow long range fluctuations with increasing frequencies.    

The Crossing function can have different forms, but considering the shape of the data and our expectations that the angular power 
spectrum should exhibit smooth behavior in the context of the concordance model, we expect Chebyshev polynomials of the first kind will perform satisfactorily. 
In this regard one can find some detailed discussions in ~\cite{Shafieloo:2012jb}.


It will be explained in the results section that we use Chebyshev polynomials only up to fourth order. Due to convergence of the 
results there is usually no need to go to the higher orders. One can use higher orders of Chebyshev polynomials but this would introduce more degrees of freedom,  
larger confidence limits that eventually restrict us from distinguishing between cases. \\

3. Now we fit $C_{\ell}^{\rm TT}\mid_{\rm modified}^{N}$ to WMAP 9 year data and derive the confidence limits of the Crossing hyperparameters $C_0,C_1,C_2,...,C_N$. These
coefficients of the Chebyshev polynomials (Crossing hyperparameters), perform as discriminators between true and false models. Eq.\ref{eq:mainmain} generates
many sample models, all based on the given mean function and each $C_1,C_2,...,C_N$ set produces variation with a particular likelihood fitting the data. This likelihood represents 
the ``probability of a particular variation given the data''. $1\sigma$ and $2\sigma$ contours represent $68\%$ and $95\%$ confidence limits. As an example, in case of considering $T_{II}$,
any point (in the marginalized $C_1-C_2$ hyperparameter space) within $2\sigma$ probability contour, represents a variation that given this variation, the probability of the observed data
would be more than $5\%$. Now, depending on where the $C_0=1, C_{1-N}=0$ point stands (which is the 
variation associated to the mean function with no change) in comparison to the $N-\rm Dimensional$ confidence ball, one can state whether the given mean function can express the data 
to a given significance. Let us recall that this mean function is the best fit model to the other dataset. \\

As we have stated previously $C_{\ell}^{\rm TT}\mid_{\rm best~ fit~ model}$ is the best fit TT angular power spectra obtained from fitting a particular dataset (say, Planck or WMAP). 
For a particular dataset this mean function is fixed as the parameters $\Omega_{\rm b},~\Omega_{\rm CDM},~H_0$ and $\tau$ along with $A_{\rm S}$ and $n_{\rm S}$ are fixed to their best
fit values~\footnote{We have calculated the mean functions using CAMB~\cite{cambsite,Lewis:1999bs}}. While fitting the $ C_{\ell}^{\rm TT}\mid_{\rm modified}$ to the other dataset we perform 
Markov chain Monte Carlo (MCMC) analysis to put constraints on the Crossing hyperparameters. 
For MCMC we have used the publicly available software CosmoMC~\cite{cosmomcsite,Lewis:2002ah}. We should note that in our comparison between the two datasets we have kept the polarization 
sector untouched as Planck has not yet published its polarization data and our analysis is 
based on the temperature data from WMAP-9 and Planck. Since the reported best fit models from WMAP-9 and Planck both include WMAP-9 low-$\ell$ (up to $\ell=23$) polarization data, we have
used same best fit models as our mean functions but beyond that we have only 
considered the temperature data from the two observations for our likelihood analysis.

In this paper, we examine a two-way consistency check. First we use the Planck 
best fit spectrum as the mean function and compare with the WMAP-9 data, which clarifies whether WMAP-9 data is consistent with Planck best fit model. Then we consider the 
WMAP 9 year best fit model as the mean function and compare it with Planck data to check the reverse (whether the best fit WMAP 9 year model is consistent to Planck data). It should 
be mentioned that the two way consistency check is important as if the best fit model from WMAP is consistent with Planck data, it does not necessarily imply that the best fit 
model from Planck is consistent with WMAP-9 data. Planck has much higher precision than WMAP and has provided the data to much higher multipoles ($\ell=2500$). 
Hence a theoretical model that fits Planck data may also fit WMAP-9 data (if there is no systematic) since WMAP-9 data span over a smaller range $\ell<1200$, but 
the converse may not be true. 
This can be easily understood if we consider the different volumes of the N-Dimensional confidence balls 
(N is associated to degrees of freedom of the assumed modified model) when we fit a particular modified model to two datasets with different precision. Fitting the same modified model, the 
volume of the confidence
ball for the data with higher precision would be smaller than the volume of the confidence ball for the data with lower precision. Hence, a point in the smaller confidence ball associated 
to the data with higher precision would be included in the confidence ball of the data with lower precision but the reverse may not necessarily be true. Overall one can argue that it is always 
better to choose the mean 
function from the data with higher precision and broader range since the interpretation of the results would be straightforward as a 
mean function derived from the broader and more precise data covers the whole range of the other dataset.

We would like to note here that in a same line of our study one could possibly use some other approaches too to compare the two data sets. 
For example Gaussian processes can be used to model the two data sets using a single mean function and compare the 
confidence limits of the hyperparameters~\cite{Rasmussen,Shafieloo:2012ms}. Information field theory with 
some adjustments can be possibly used as well~\cite{Caticha,Ensslin:2013ji}.

Before going to the results section we should state that when we compare the modified models with WMAP-9 data we have included the Sunyaev-Zeldovich effect
and in comparing our modified models with the Planck data we explored the parameter space of the foreground effects since the foregrounds are quite important on small 
scales. We have marginalized over 14 nuisance parameters for different foreground and calibration effects~\footnote{For more discussions see~\cite{Ade:2013zuv}.}.

  

\section{Results}\label{sec:results}

In this section we shall present our results and address the consistency of the two recent CMB datasets. To begin, we test the robustness of the method and perform a self-consistency
test. We fit the modified best fit WMAP-9 model (by modified we mean the best fit model, which is our mean function, multiplied by the Crossing function) to WMAP-9 data itself 
and assess the fiducial point in the confidence contours
\begin{figure}[!htb]
\begin{center} 
%
\resizebox{250pt}{170pt}{\includegraphics{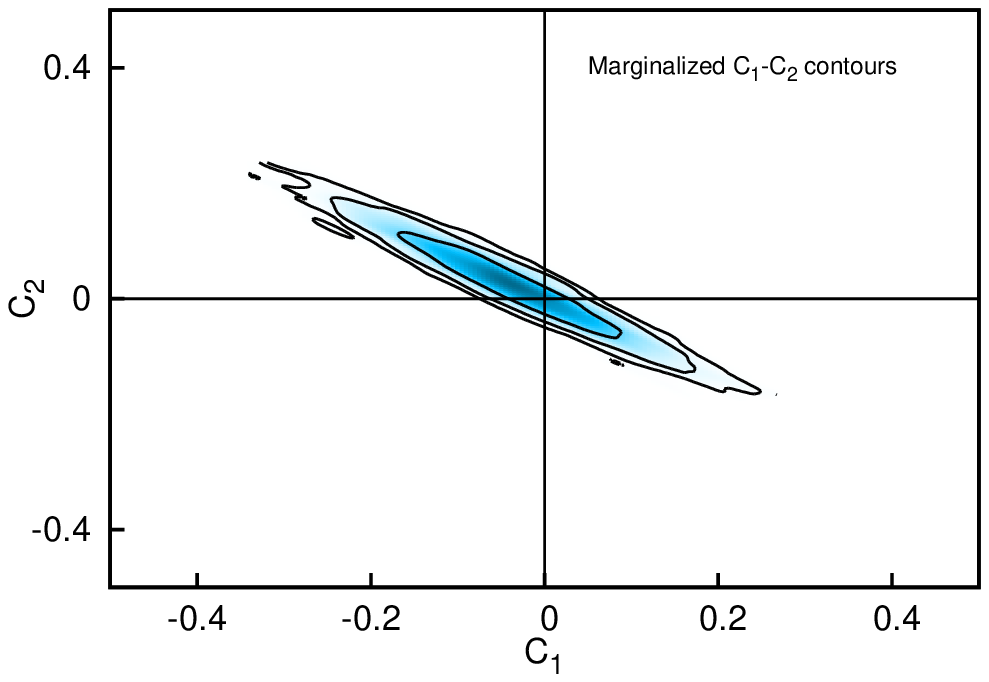}} 
\hskip -40pt \resizebox{250pt}{170pt}{\includegraphics{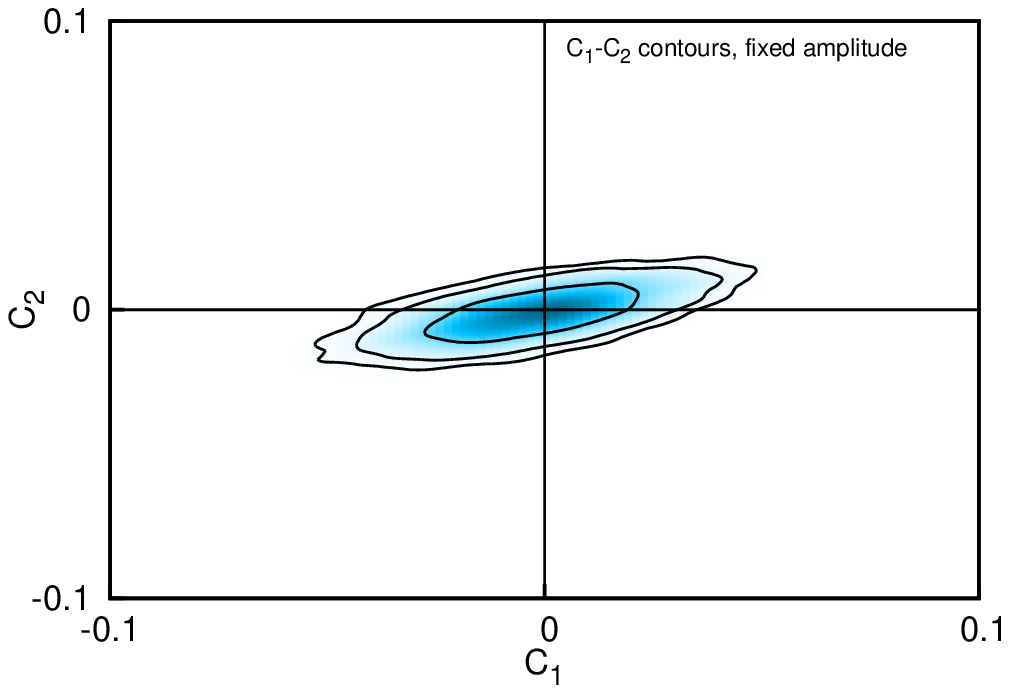}} 
\end{center}
\caption{\footnotesize\label{fig:cntrs9to9}  In this plot the 3$\sigma$ contours of the Crossing hyperparameters are plotted when we 
modify the WMAP 9 year best fit spectrum with a Crossing function (here Chebyshev polynomials) and compare with the WMAP 9 year data itself. All the plots
show that the fiducial model (corresponding and $C_{1,2}=0$) lies well inside the 1$\sigma$ contours of $C_1-C_2$. The plot in the left panel is obtained when 
we marginalized over $C_0$ and in the right panel we fixed $C_0=1$. The fiducial model remains nearly at the center of the 1$\sigma$ CL in both cases. The reduced 
size of contours in the right panel is due to the fact that when we fix the amplitude the degeneracies decrease. In this particular analysis we have fixed $\ell_{\rm max}=1200$, 
however for plots hereafter we have used $\ell_{\rm max}=2500$, the maximum multipole covered by Planck.}
\end{figure}
of the hyperparameters. In Fig.~\ref{fig:cntrs9to9} we have plotted the marginalized confidence contours of $C_{1,2}$ (left panel). It is 
evident that the mean function (denoted by the intersection of the two perpendicular black lines at the center 
of each plot) lies inside and almost at the center of the 1$\sigma$ contour. This indicates that the best fit model from WMAP-9 is 
consistent to WMAP 9 year data. It should be noted that in this case any significant deviation from mean function could indicate 
inconsistency of the concordance model to the data. This result 
is in agreement with our previous analysis on WMAP-9 data~\cite{Hazra:2013xva}. Moreover, as an 
additional test we have fixed the overall amplitude to its fiducial value ($C_0=1$) and performed MCMC on $C_1$ and $C_2$. We have plotted 
the result in the right panel of Fig.~\ref{fig:cntrs9to9}. As expected, the size of the contours decreased considerably as the degeneracies 
between the hyperparameters are lifted by fixing the overall amplitude. Here too we find that the fiducial best 
fit model lies near to the center of 1$\sigma$ limits of hyperparameters $C_1-C_2$. These figures indicate that the best fit concordance model 
to WMAP 9 year data is indeed in good agreement with the data.

 \begin{figure}[!htb]
\begin{center} 
%
\resizebox{250pt}{170pt}{\includegraphics{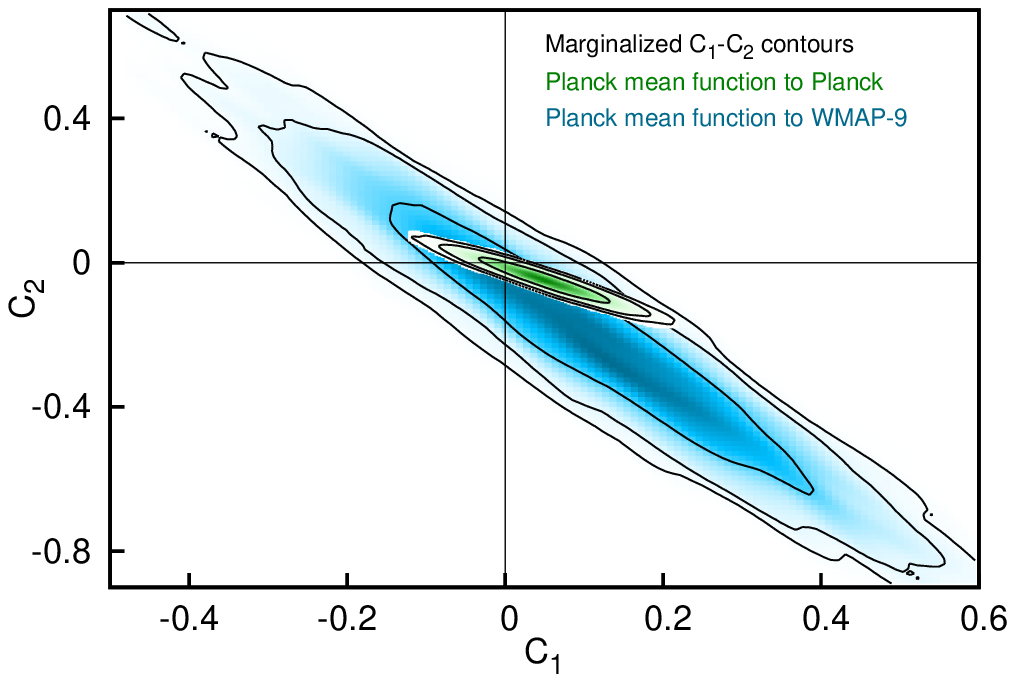}} 
\hskip -40pt\resizebox{250pt}{170pt}{\includegraphics{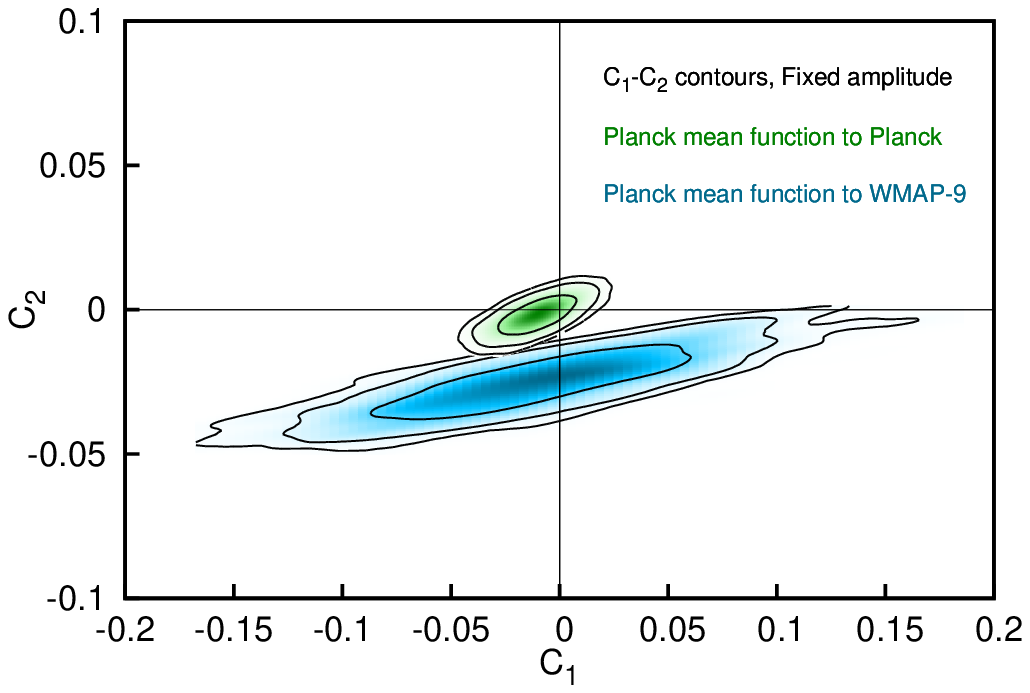}} 
\end{center}
\caption{\footnotesize\label{fig:cntrsplanckto9} Plots obtained using Planck best fit concordance model as the mean function to fit WMAP-9 and Planck data considering a Crossing function of 
the second order. We have used Chebyshev polynomials as the Crossing functions. Plots show that allowing the overall amplitude to vary (left panel) the two observations are in
good agreement with each other having a large overlap of the confidence contours at $1\sigma$ level. The second plot is obtained by keeping $C_0=1$ (fixed amplitude) and 
varying the first and second order hyperparameters. In this case we can see a clear discrepancy between the two datasets. Our results indicate that if both 
observations insist on their reported values of the overall amplitude of the angular power spectrum, the inconsistency between the two datasets would be unresolvable.
 } 
\end{figure}

 Now we fit the modified mean functions of the two datasets (best fit concordance models to WMAP-9 and Planck data multiplied by the 
Crossing function) to WMAP-9 and Planck data.  In Fig.~\ref{fig:cntrsplanckto9} we exhibit the confidence contours of the 
Crossing hyperparameters (in this case the coefficients of the Chebyshev polynomial of second order) when we fit the modified mean 
function from Planck data (best fit concordance model to Planck data multiplied by the Crossing function) to Planck and WMAP 9 year data. 
In the left panel the $C_1-C_2$ contours are marginalized over $C_0$ allowing the overall amplitudes to vary and right panel is 
for $C_0=1$, fixing the overall amplitudes at their reported values. It is interesting to see that the best fit model to 
Planck data (mean functions are denoted by the intersection of the two perpendicular black lines at the center of each plot) is consistent
with both observations (see the left panel). The large intersection area between the two confidence contours also indicates that the two observations are very 
well in agreement. One can interpret this as there being some hypothetical variations, each denoted by a point in the hyperparameter space, 
that can explain both observations simultaneously with high probabilities if we allow the overall amplitudes to vary. From the right panel of Fig.~\ref{fig:cntrsplanckto9}, 
it is evident that by fixing the amplitudes of the angular power spectra to their reported survey, 
there is a significant inconsistency - more than 3$\sigma$. This result suggests 
that by fixing the amplitudes, there is no hypothetical model that can simultaneously express both observations with high certainty. Assuming that the proposed model($\Lambda$CDM) is correct, 
this indicates that there is a clear tension between the two datasets. We conclude that the overall 
amplitudes of the angular power spectra from Planck and WMAP play a very important role to test if these datasets are consistent. If we allow these 
amplitudes to vary, there is no tension, but if both observations insist on their reported values of the overall amplitudes there will 
be an unresolvable tension between the two.               

\begin{figure*}[!htb]
\begin{center} 
\resizebox{420pt}{310pt}{\includegraphics{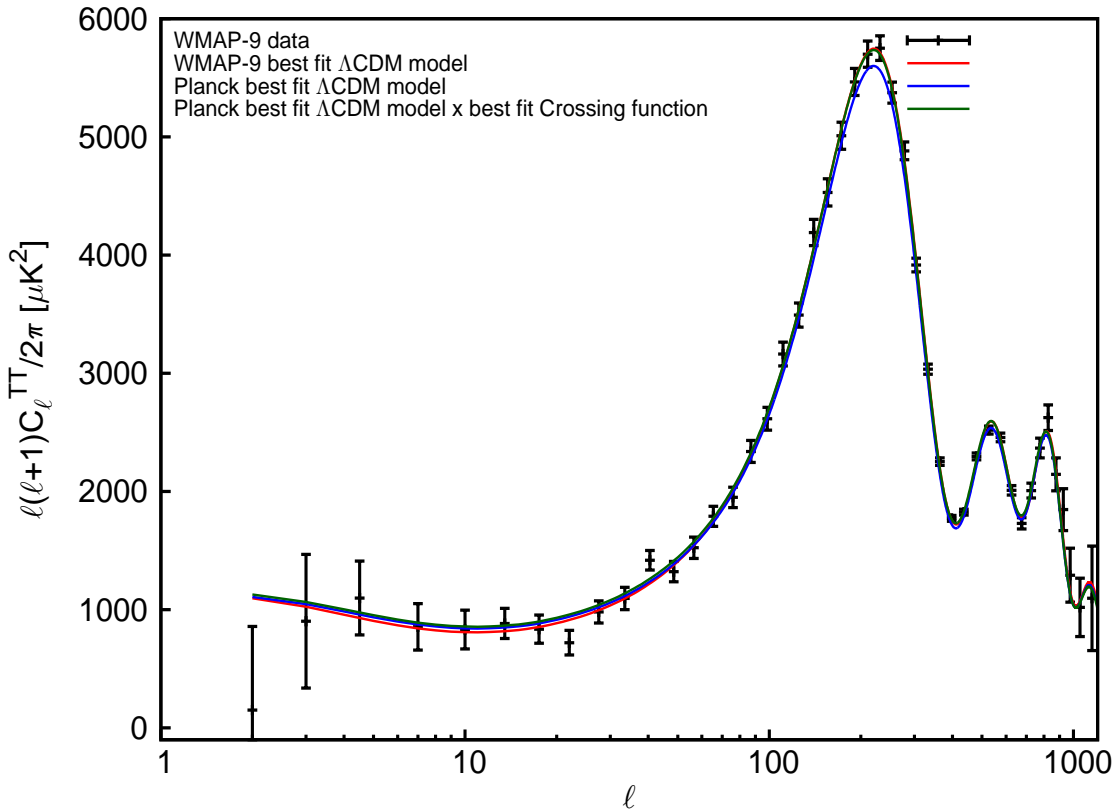}} 
\end{center}
\caption{\footnotesize\label{fig:clsto9} WMAP 9 year data for temperature auto-correlation is plotted with different theoretical $\cl$'s. The WMAP best fit 
is plotted in red while the Planck best fit is plotted in
blue. Note that the mismatch in power is evident near the first acoustic peak. It is shown that we can fit the WMAP data pretty well with Planck best model 
modified by the Crossing function (in green).}
\end{figure*}

To check the extent of the mismatch we have compared the Planck best fit model with WMAP-9 data varying only the amplitude $C_0$. The best fit 
value of $C_0$ in this case is 1.024 which implies the WMAP-9 spectrum is about $2.4\%$ higher than Planck~\footnote{This amplitude difference was mentioned 
in the Planck analysis in CMB power spectrum and likelihood~\cite{Planck:2013kta}}. Performing our Crossing analysis once 
again fixing $C_0=1.024$ we found the that the mean function point $C_{1,2}=0$ lies now completely inside the 1$\sigma$ contours of $C_1-C_2$. 

 \begin{figure*}[!htb]
\begin{center} 
%
\resizebox{250pt}{170pt}{\includegraphics{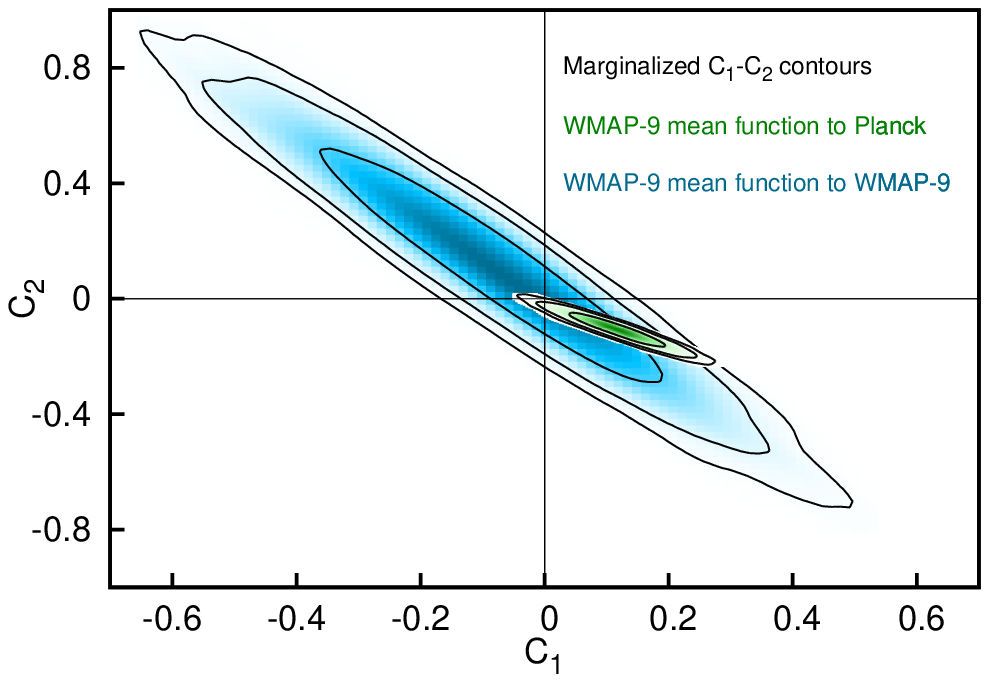}} 
\hskip -40pt\resizebox{250pt}{170pt}{\includegraphics{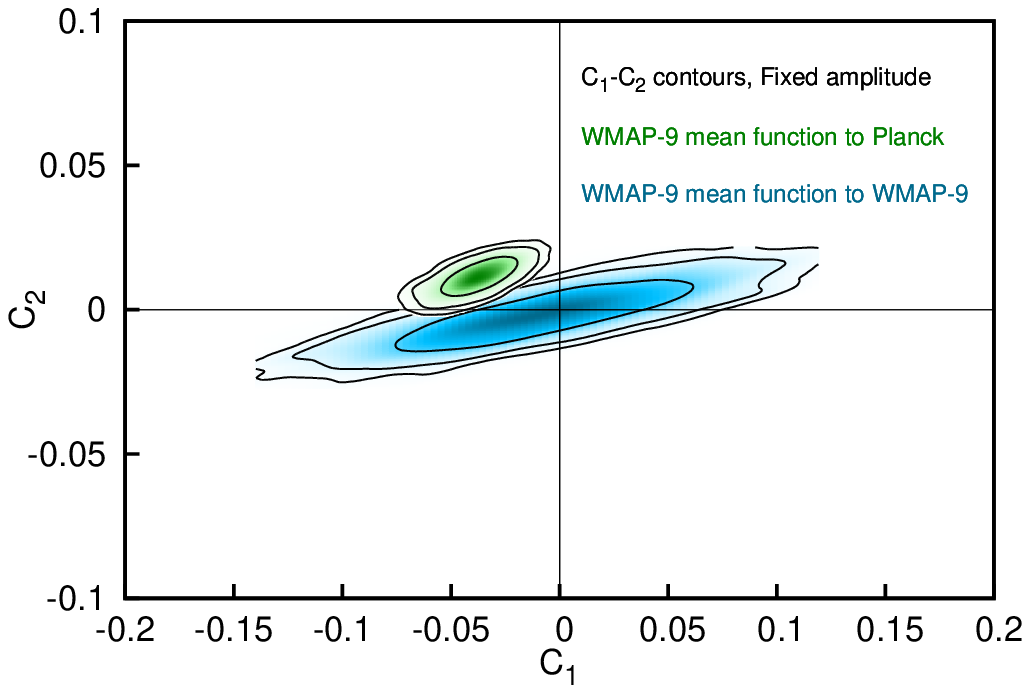}} 

\end{center}
\caption{\footnotesize\label{fig:cntrs9toplanck} Plots obtained using WMAP 9 year best fit spectrum as the mean function to fit WMAP-9 and Planck data considering a Crossing 
function of the second order. We have used Chebyshev polynomials as the Crossing functions. Plots clearly dictate that the WMAP-9 best best fit model is not consistent with Planck data. 
When we fix the overall amplitude $C_0=1$, the fiducial model becomes further away from the confidence contours (right panel). Looking at these results still we 
cannot conclude that the two datasets are inconsistent. In fact the proper overlap of the confidence contours in the left panel (allowing a shift in the overall amplitude) 
reflect the consistency of the two datasets.   
}
\end{figure*}

 In Fig.~\ref{fig:clsto9} we show the WMAP 9 year binned TT spectrum with some theoretical $\cl$'s overplotted. Specifically we show 
 the best fit theoretical model from WMAP-9 (red) and Planck (blue). These two curves clearly show that WMAP power is higher than Planck in 
 high-$\ell$ and this is most evident near the first acoustic peak. 
 However, the green curve, the Planck best fit modified with best fit Crossing function, shows that difference
 between Planck and WMAP-9 data can be addressed using a second order Crossing function.

We should mention here that a Crossing function allows us to match the best fit models from the two datasets up to a statistically indistinguishable 
level (depending on the quality of the data) using an orthogonal basis independent of the cosmological parameters. This in turn allows us to understand  
how far these models are from each other statistically and if there is any hypothetical variation that can explain both datasets simultaneously.

Having described the consistency of Planck and WMAP-9 data using the best fit model from Planck as the mean function, we shall demonstrate now 
our results 
when we use the best concordance model to WMAP 9 year data as the mean function. We should note here that comparing two datasets, it is always better to 
choose a mean function from the data that has a higher precision and covers a broader range of scales. 
Despite this, it is still of interest to perform the reverse test. 
We start with 
a modulation by Chebyshev polynomials up to second order. We should emphasize that here we allow the 14 nuisance parameters for foreground and calibration to 
vary along with the Crossing hyperparameters. The contours in Fig.~\ref{fig:cntrs9toplanck} show that the WMAP-9 mean function is 3$\sigma$ away in the 
$C_1-C_2$ contour plane fitting Planck data. When we fix the overall amplitude $C_0=1$ we find the smaller contour fitting the Planck data pushes the mean function point further away. 
This result clearly indicates that the best fit concordance model to WMAP 9 year data is inconsistent at not less than 3$\sigma$ level from Planck data. This result is completely 
consistent with the previous analysis.

As we have noted earlier, when fitting the same modified model, the area of the marginalized confidence contours (of the hyperparameters) for the data with higher precision would 
be smaller than the  marginalized confidence contours for the data with lower precision. Hence, even if both data sets are different realizations of the same 
fiducial model (and by default consistent), 
a point in the smaller confidence contour associated to the data with higher precision would be necessarily included in the confidence contour of the data with lower precision but the 
reverse may not be always true. If we look at the plots in Fig.~\ref{fig:cntrsplanckto9} and Fig.~\ref{fig:cntrs9toplanck} it is evident that the contours fitting Planck data is 
much smaller than the ones fitting WMAP-9. This is simply due to the higher precision and broader range of the Planck data.

\begin{figure*}[!htb]
\begin{center} 

\resizebox{420pt}{310pt}{\includegraphics{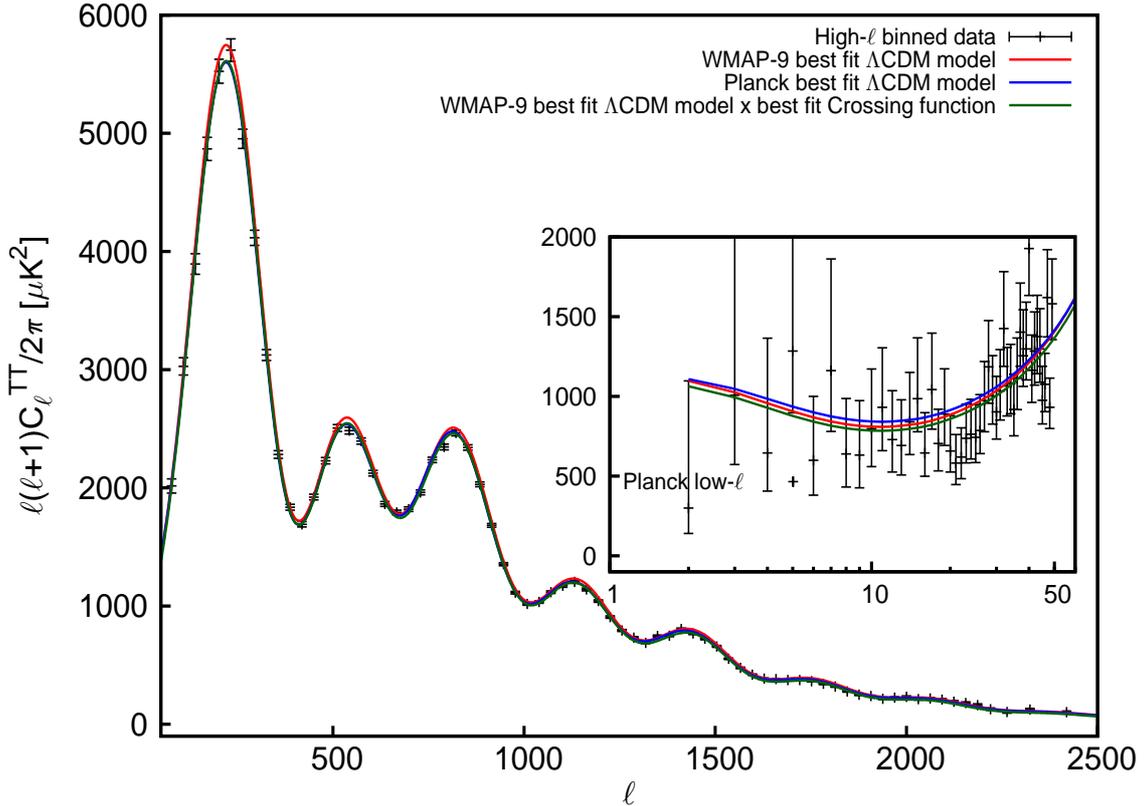}} 
\end{center}
\caption{\footnotesize\label{fig:clstoplanck} Planck data and various theoretical $\cl$'s are plotted here. The color codes of WMAP-9 and 
Planck best fit are the same as in Fig.~\ref{fig:clsto9}. The green line corresponds
to the WMAP-9 best fit model modified with best fit Crossing function. It can be clearly seen that the amplitude difference is appropriately addressed 
with the Crossing function. In the inset, Planck data for low-$\ell$ ($\ell=2-49$) is plotted. High-$\ell$ binned spectrum is plotted in the main plot
which provides an idea of the differences between Planck and WMAP mean functions and the modified spectrum.}
\end{figure*}

\begin{figure*}[!htb]
\begin{center} 
\includegraphics[angle=270,width=110mm]{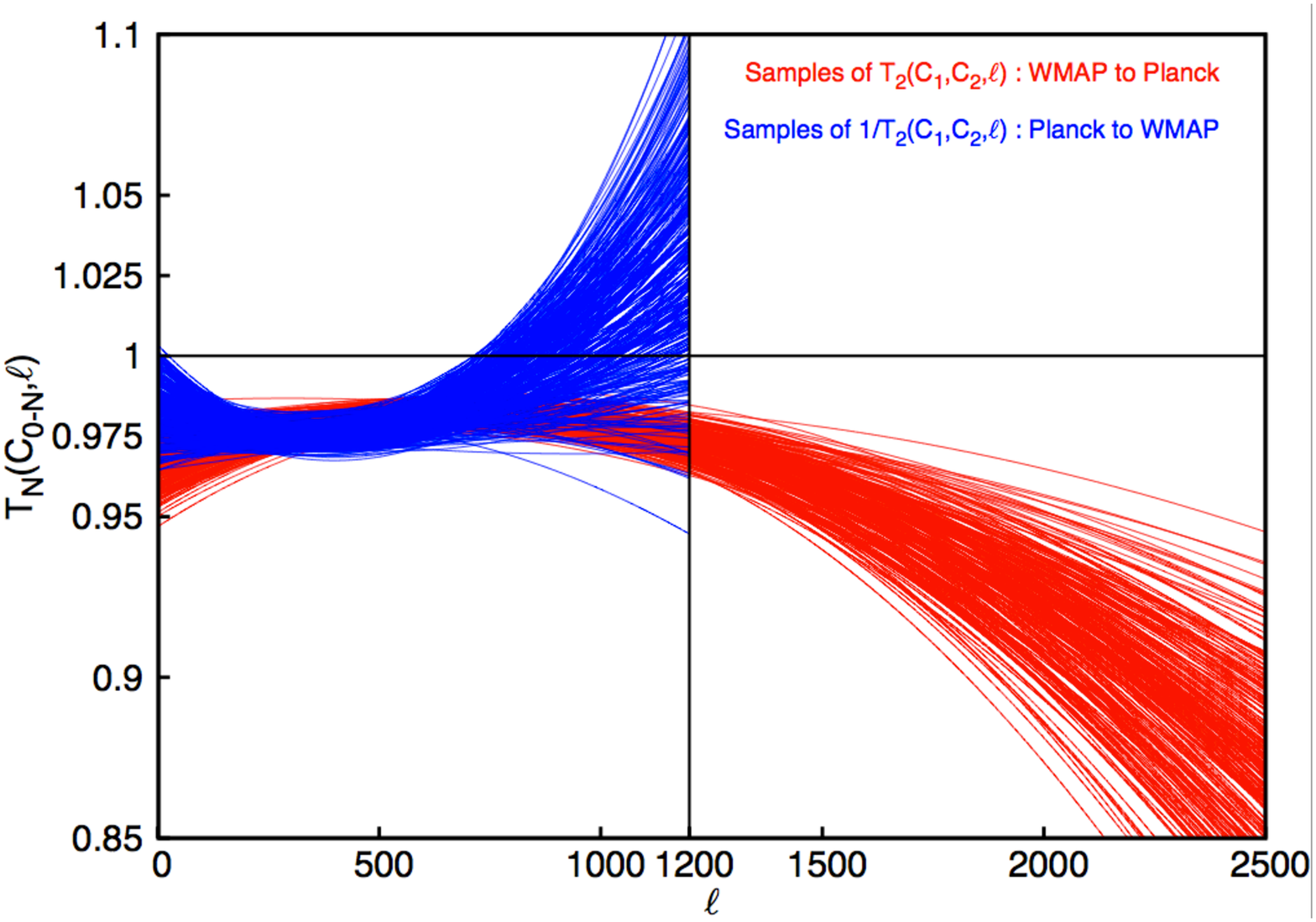}
\end{center}
\vskip -15pt
\caption{\footnotesize\label{fig:cf} Samples of Crossing and inverse Crossing functions within the 2$\sigma$ allowed 
range of Crossing hyperparameters are plotted. The largest multipole range covered by WMAP is indicated by the vertical line.
}
\end{figure*}

Using the second order Chebyshev polynomials to modify the WMAP-9 best fit model, we have not yet reached the likelihood which is 
comparable to Planck best fit model (though our analysis was sufficient to show that the mean function from WMAP is not consistent to Planck data). Hence we now modify 
the spectrum further using the third and the fourth order Chebyshev polynomials. 
While the best fit WMAP-9 model had $\Delta\chi^2\sim80$ worse with respect to the best fit Planck model to Planck data, a significant difference, when we use second and third order 
Crossing modifications the $\Delta\chi^2$ reduces to 16 and 7 respectively and with fourth order it reduces further to less than 5. This indicates that by
 assuming fourth order Chebyshev polynomials we are able to almost match the best fit model from WMAP to the best fit model from Planck. In Fig.~\ref{fig:clstoplanck} 
 we have plotted the Planck data along with some theoretical $\cl$'s overlayed. The red and the blue lines are the 
 WMAP-9 and Planck best fit models respectively. The green curve is for the modified WMAP-9 best fit model with Crossing function 
 (to second order) set to its best fit form. It is clear that with the Crossing function, we can address the WMAP-9 and Planck discrepancies. 
 The Planck high-$\ell$ binned spectrum is shown 
in the main plot and the low-$\ell$ data is plotted in the inset. It is clearly visible that the modified WMAP-9 spectrum matches the 
Planck best fit at all scales.

In figure~\ref{fig:cf} we have plotted the samples of Crossing functions within 2$\sigma$ confidence contours of the Crossing 
hyperparameters. Through the red lines we have shown the samples of second order Crossing function when modified mean function from WMAP-9 is compared with Planck data 
and in blue lines we have compared the former with the inverse Crossing functions (to second order) when the modified mean function from Planck is compared with WMAP-9 data. 
The smallest scale probed by WMAP ($\ell=1200$) is indicated by a vertical line. Apart from an amplitude shift the plots successfully capture 
the scale dependent mismatch between WMAP and Planck. The agreement of the Crossing functions and the inverse Crossing functions in this plot 
reflects the robustness of our analysis. Note that the blue curves contain a few lines which are nearly scale invariant, signifying 
that allowing for an overall amplitude shift the Planck mean function can address the WMAP-9 data very well without further modification. This agreement basically reflects
the results in the left panel of Fig.~\ref{fig:cntrsplanckto9}. 
However, most importantly it can be seen that the WMAP mean function requires further modifications than just an overall amplitude shift to address the Planck data (the red 
 curves), especially, the preferred modifications (beyond the overall amplitude shift) are more evident in the smaller scales probed by Planck ($\ell>1200$). For the range of scales 
 probed by WMAP ($\ell=2-1200$) the Crossing functions suggest an overall amplitude correction.

\section{Discussion}\label{sec:discussion}

In this paper, which is the first of the two consistency check papers, we have addressed the 
consistency of WMAP-9 data to the recent Planck dataset within the framework of the concordance 
$\Lambda$CDM model. In the context of Crossing statistic, we adopted the best fit concordance model 
to WMAP 9 year (Planck) data as the mean function and multiplied it to a Crossing function to fit 
Planck (WMAP 9 year) data. The marginalized confidence contours of the Crossing hyperparameters have been 
used to check the consistency of the two observations. In our analysis we used Chebyshev polynomials 
as the Crossing functions. Allowing the overall amplitude of the datasets to vary, which is an 
important assumption, we found that the best fit model to Planck data is
consistent with WMAP 9 year data. However, contrary to this, we found that the best fit 
model to WMAP-9 data is not consistent to Planck data at worse than $3\sigma$. This is
not particularly unexpected since the precision of the Planck data is considerably higher than 
WMAP-9 data. Since 
the area of the confidence contours of the hyperparameters using Planck data are smaller than the area of the confidence 
contours using WMAP 9 year data, not all points in the confidence contours of WMAP-9 (models consistent to WMAP-9) 
should remain within smaller confidence contours of Planck. Looking at the large overlap of the confidence contours of the 
Crossing hyperparameters (Fig.~\ref{fig:cntrsplanckto9}) we can conclude that allowing the overall shift of the amplitude of the 
two spectra, there is 
no strong statistical evidence that Planck and WMAP-9 data are inconsistent. Our results show that there are some
hypothetical models from which both data can be simultaneously drawn. 
In other words, there are some hypothetical models that given these variations, the probabilities of both observed data are high. 
In the plot of the Crossing functions we have shown a family of modifications which can explain the possible discrepancies 
between the two datasets as a function of angular scales apart from an overall amplitude factor.

This highlights the fact that comparing the 
derived individual cosmological parameters from different data sets might be misleading when the observables
are complicated functions of these parameters. The basic six cosmological parameters of the concordance model 
have a very different nonlinear effect on the angular power spectrum and we use Boltzmann codes to derive $C_{\ell}$'s. 
These parameters are degenerate and they are not orthogonal to each other. However, the Chebyshev 
polynomials that we used as our Crossing functions have properties of 
orthogonality which makes it suitable to test the consistency of the data from two observations. 

We should emphasize that the consistency of Planck and WMAP 9 year data holds only if we allow the overall
amplitudes of the two spectra to vary. Fixing the amplitudes at their reported values results in a clear and significant 
inconsistency of the two data sets at more than $3\sigma$ and the best fit models
from both observations are ruled out strongly by the other observation. In this case, in fact there is no hypothetical variation that can express both datasets simultaneously
with high probabilities. This is an important issue to be resolved. Resolving this issue, 
one can claim that the two observations are consistent.


\section*{Acknowledgments}
D.K.H and A.S wish to acknowledge support from the Korea Ministry of Education, Science
and Technology, Gyeongsangbuk-Do and Pohang City for Independent Junior Research Groups at 
the Asia Pacific Center for Theoretical Physics. A.S would like to acknowledge the support of the National Research Foundation of Korea (NRF-2013R1A1A2013795). We also acknowledge the use of publicly 
available CosmoMC in our analysis. D.K.H and A.S would like to thank Eric Linder and Tarun Souradeep
for useful discussions. We thank Stephen A. Appleby for his useful comments on the manuscript. 
We acknowledge the use of WMAP-9 data and likelihood from Legacy Archive for 
Microwave Background Data Analysis (LAMBDA)~\cite{lambdasite} and Planck 
data and likelihood from Planck Legacy Archive (PLA)~\cite{PLA}.


\end{document}